\documentclass[prb,twocolumn,showpacs,amsmath,superscriptaddress,psfig,amssymb]{revtex4-1}

\usepackage{amssymb,color}
\usepackage{graphicx}
\usepackage{dcolumn}
\usepackage{bm}
\usepackage[utf8]{inputenc}
\usepackage{mathtools}

\usepackage{color}

\usepackage[pdfstartview=FitH, CJKbookmarks=true, bookmarksnumbered=true, bookmarksopen=true, colorlinks, pdfborder=001, linkcolor=blue, anchorcolor=blue, citecolor=blue]{hyperref}

\begin{document}

\title{Enhancement of the effective mass at high magnetic fields in CeRhIn$_5$}

\author{L. Jiao}
\affiliation{Center for Correlated Matter and Department of Physics, Zhejiang University, Hangzhou, Zhejiang 310058, China}
\affiliation{Max Planck Institute for Chemical Physics of Solids, 01187 Dresden, Germany}
\author{M. Smidman}
\email{msmidman@zju.edu.cn}
\affiliation{Center for Correlated Matter and Department of Physics, Zhejiang University, Hangzhou, Zhejiang 310058, China}
\author{Y. Kohama}
\affiliation{Institute for Solid State Physics, The University of Tokyo, Kashiwa, Chiba 277-8581, Japan}
\author{Z. S. Wang}
\altaffiliation{Present address: High Magnetic Field Laboratory of the Chinese Academy of Sciences, Hefei 230031, China}
\affiliation{Hochfeld-Magnetlabor Dresden (HLD-EMFL), Helmholtz-Zentrum Dresden-Rossendorf, D-01328 Dresden, Germany}
\author{D.~Graf}
\affiliation{National High Magnetic Field Laboratory, Florida State University, Tallahassee, FL 32310}
\author{Z.~F.~Weng}
\affiliation{Center for Correlated Matter and Department of Physics, Zhejiang University, Hangzhou, Zhejiang 310058, China}
\author{Y.~J.~Zhang}
\affiliation{Center for Correlated Matter and Department of Physics, Zhejiang University, Hangzhou, Zhejiang 310058, China}
\author{A.~Matsuo}
\affiliation{Institute for Solid State Physics, The University of Tokyo, Kashiwa, Chiba 277-8581, Japan}
\author{E.~D.~Bauer}
\affiliation{Condensed Matter and Magnet Science, Los Alamos National Laboratory, Los Alamos, NM, USA}
\author{Hanoh Lee}
\affiliation{Center for Correlated Matter and Department of Physics, Zhejiang University, Hangzhou, Zhejiang 310058, China}
\author{S. Kirchner}
\affiliation{Zhejiang Institute of Modern Physics, Zhejiang University, Hangzhou, Zhejiang 310027, China}
\author{J.~Singleton}
\affiliation{Condensed Matter and Magnet Science, Los Alamos National Laboratory, Los Alamos 87545, NM, USA}
\author{K.~Kindo}
\affiliation{Institute for Solid State Physics, The University of Tokyo, Kashiwa, Chiba 277-8581, Japan}
\author{J.~Wosnitza}
\affiliation{Hochfeld-Magnetlabor Dresden (HLD-EMFL), Helmholtz-Zentrum Dresden-Rossendorf, 01328 Dresden, Germany}
\author{F.~Steglich}
\affiliation{Center for Correlated Matter and Department of Physics, Zhejiang University, Hangzhou, Zhejiang 310058, China}
\affiliation{Max Planck Institute for Chemical Physics of Solids, 01187 Dresden, Germany}
\author{J.~D.~Thompson}
\affiliation{Condensed Matter and Magnet Science, Los Alamos National Laboratory, Los Alamos 210093, NM, USA}
\author{H.~Q.~Yuan}
\email{hqyuan@zju.edu.cn}
\affiliation{Center for Correlated Matter and Department of Physics, Zhejiang University, Hangzhou, Zhejiang 310058, China}
\affiliation{Collaborative Innovation Center of Advanced Microstructures, Nanjing University, Nanjing, China}

\begin{abstract}
The Kondo-lattice compound CeRhIn$_5$ displays a field-induced Fermi surface reconstruction at $B^*\approx30$~T, which occurs within the antiferromagnetic state, prior to the quantum critical point at $B_{c0}\approx50$~T. Here, in order to investigate the nature of the Fermi surface change, we measured the magnetostriction, specific heat, and magnetic torque of CeRhIn$_5$ across a wide range of magnetic fields.  Our observations uncover the field-induced itineracy of the $4f$ electrons, where above $B_{\rm onset}\approx17$~T there is a significant enhancement of the Sommerfeld coefficient, and spin-dependent effective cyclotron masses determined from quantum oscillations. Upon crossing $B_{\rm onset}$,  the temperature dependence of the specific heat also shows distinctly different behavior from that at low fields. Our results indicate that the Kondo coupling is remarkably robust upon increasing the magnetic field. This is ascribed to the delocalization of the $4f$ electrons at the Fermi surface reconstruction at  $B^*$.
\end{abstract}


\maketitle
\section{introduction}
Kondo-lattice systems are prototypical strongly correlated
materials, in which thermal and quantum fluctuations can drive the electronic
states in spin, charge, or orbital channels and induce
various phases. In the simplest model, the ground states of  heavy-fermion
compounds are determined by the competition between the Ruderman-Kittel-Kasuya-Yosida (RKKY)
interaction and the Kondo effect, which evolve differently with
temperature, magnetic field, and pressure \cite{Stewart,Hilbert,Gegenwart,Weng2016}. Adjusting  non-thermal control parameters can tune the strengths of both the RKKY coupling, which is the intersite exchange interaction between localized moments leading to magnetic order, and the Kondo coupling corresponding to onsite antiferromagnetic exchange  between localized moments and itinerant conduction electrons, giving rise to a nonmagnetic singlet ground state. To date, many novel phenomena, such as non-Fermi-liquid behavior and unconventional
superconductivity, have been found in the vicinity of quantum
critical points (QCPs), where the Kondo coupling prevails and the magnetic ordering temperature is smoothly suppressed to zero. However, whether a universal description of quantum critical behavior can be found still needs to be determined, and in particular it is necessary to investigate  how the emergent phenomena evolve upon tuning Kondo lattice systems from localized to itinerant states.

The Kondo lattice compound CeRhIn$_5$  provides a good opportunity to
study the change from localized to delocalized $4f$ states. In zero-field at ambient pressure, the Ce 4$f$
electrons in CeRhIn$_5$ are localized and order antiferromagnetically at around
3.8~K \cite{Hegger,Shishido2002,Harrison2004}. In high magnetic fields, evidence for a field-induced
spin-density-wave (SDW) type QCP has been found near the critical field
$B_{c0}\approx50$~T \cite{Jiao}.
Theoretically, this kind of conventional SDW-type QCP involves only three-dimensional fluctuations of the
antiferromagnetic (AFM) order parameter around zero temperature
\cite{Hertz,Moriya,Millis}, which has been identified in some heavy-fermion compounds,
such as CeCu$_2$Si$_2$ \cite{Arndt}, and La-doped CeRu$_2$Si$_2$ \cite{Kambe}.
In this scenario, the $f$ electrons already are delocalized in the magnetically ordered
region, and only a few ``hot spots/lines" on the Fermi surface (FS) become critical
at the QCP \cite{Gegenwart}. For CeRhIn$_5$, a field-induced change of the FS was
observed inside the AFM state at around $B^{\ast}$$\approx$~30~T, as concluded
from sharp changes in the dHvA frequencies, Hall effect, and magnetoresistance
\cite{Jiao,Moll,Jiao2}. The expansion of the FS volume (or charge carrier concentration) is attributed
to the $4f$ electrons becoming delocalized at $B>B^*$ \cite{Jiao}, which is consistent with the presence of a  conventional SDW-type QCP at  $B_{c0}$. Such a field-induced delocalization of the $4f$ electrons is different to that generally anticipated for Kondo lattice systems, where the $4f$ electrons are often expected to be more localized in high fields \cite{Aoki1993,Harrison1993,Harrison1998,Ce142}.

To date, no clear anomalies have been detected around $B^*$ in the specific heat \cite{Jiao}, torque
magnetometery \cite{Jiao,Moll}, and magnetic susceptibility \cite{Takeuchi}. Therefore, the evolution of the
Ce 4$f$ electrons in high magnetic fields is still not determined, and the nature of
the quantum phase transition at $B^*$ is not well understood \cite{Jiao}. Moreover, evidence was recently provided
for a spin nematic state in CeRhIn$_5$ at high fields, from resistivity measurements of microstructured devices for certain field orientations \cite{Ronning2017}. A pronounced in-plane
anisotropy was found in the resistivity when a field was applied 20$^\circ$ away from the $c$~axis, where
the anisotropy showed a sudden increase above $B^*$. Obviously, the relationship between this enigmatic
phase transition and the local to itinerant transition of the $4f$ electrons needs to be resolved.
Here, we report a detailed study of the magnetostriction, specific heat, and de Haas-van
Alphen (dHvA) effect of CeRhIn$_5$ in high magnetic fields. These results enable us
to trace the evolution of the correlated electrons upon approaching $B^*$.

\section{Experimental details}

Single crystals of CeRhIn$_5$ were grown using In flux, as described elsewhere \cite{Jiao}.
Measurements of the absolute values of the specific heat are extremely challenging in pulsed
magnetic fields. These were performed in a long-pulsed magnet at the
International Megagauss Science Laboratory of ISSP in Kashiwa up to 43.5~T. To
minimize the influence associated with the field instability of the pulsed field,
we generated highly stabilized ($\pm$100 Oe) magnetic fields with a 100 ms timescale
using a field-feedback controller \cite{Yoshi1} and measured the absolute value of
the specific heat in the stabilized field by applying the heat-pulse method \cite{Yoshi2}.
The resolution of the specific heat data were successfully improved \cite{sup}, and a
detailed analysis of the data is possible up to 43.5~T. The optical fibre Bragg
grating (FBG) method \cite{Daou,Rotter} was used to measure the magnetostriction in
pulsed magnetic fields up to 60 T with $B$~$\parallel$~$a$ and $B$~$\parallel$~$c$.
Measurements of the dHvA effect of CeRhIn$_5$ were conducted utilizing a
torque technique between 5~K and 300~mK in dc fields up to 45~T \cite{Jiao}.

\section{Results}
\subsection{High-field magnetostriction}

\begin{figure}[tb]\centering
 \includegraphics[width=0.85\columnwidth]{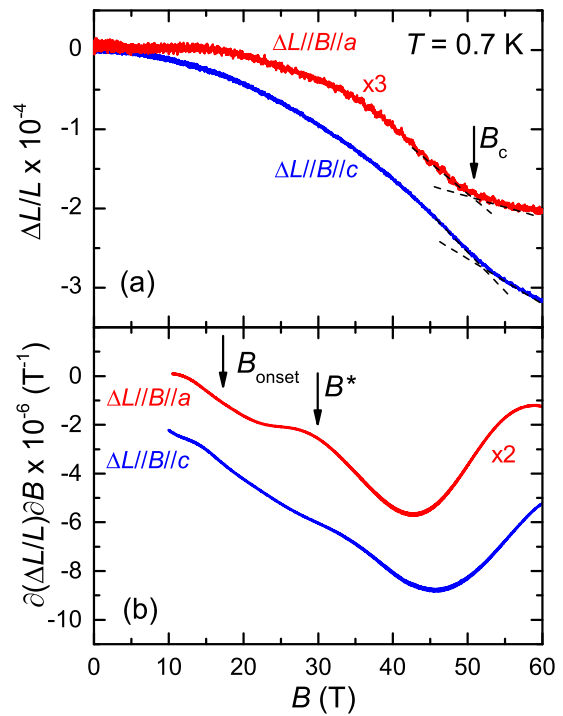}
\caption{Magnetostriction  of CeRhIn$_5$ in high magnetic fields, where (a) the fractional length change $\Delta$$L$/$L$, and (b) the derivative are displayed for $B$~$\parallel$~$a$ and $B$~$\parallel$~$c$ at 0.7~K, and  the former data have been scaled. The arrows mark the onset of the mass enhancement $B_{\rm onset}$, $B^*$ and the critical field $B_c$.}
\label{Fig1}
\end{figure}

The magnetostriction of a metal is proportional to the
conduction-electron density of states at the Fermi level \cite{Ekreem}. Therefore,  so as to gain insights into the interactions between the
conduction electrons and the lattice, we performed measurements of the  longitudinal magnetostriction of
CeRhIn$_5$ for $B$~$\parallel$~$a$ and $B$~$\parallel$~$c$ at 0.7~K,  where the relative length changes $\Delta L/L$ and the corresponding field derivatives as a function of applied field are displayed in Fig.~\ref{Fig1}. It can be seen that at lower fields up to around 15~T, the in-plane $\Delta L/L$  is nearly field independent, before beginning to decrease more rapidly at higher fields. Meanwhile at very high fields in the vicinity of the critical field $B_c$, $\Delta L/L$ shows a change in slope at 50 and 52~T for fields along the $a$- and $c$-axes respectively. While no clear anomaly is observed near $B^*\approx30$~T along the $c$~axis, there is a kink in $\partial(\Delta L/L)/\partial B$ along the $a$-axis [Fig.\ref{Fig1}(b)].  These results give further evidence for the existence of $B^*$ from in-plane measurements, but also suggest that the
coupling to the lattice of the Fermi surface transition $B^*$ is rather weak. While the critical fluctuations associated with  phenomena such as CDW transitions and valence
changes generally couple strongly to the lattice, those corresponding to SDW order commonly couple more weakly \cite{Stockert2012}. We note that weak anomalies have also been reported in the magnetostriction at $B^*$ when fields are applied at 11$^{\circ}$ and 20$^{\circ}$ to the $c$-axis \cite{Ronning2017,Rosa2018}.  It is possible that the magnetostriction change at $B^*$ is too weak to be resolved in our measurements with fields along the $c$-axis.

\subsection{Specific heat  in high magnetic fields}

\begin{figure}[tb]\centering
 \includegraphics[width=0.99\columnwidth]{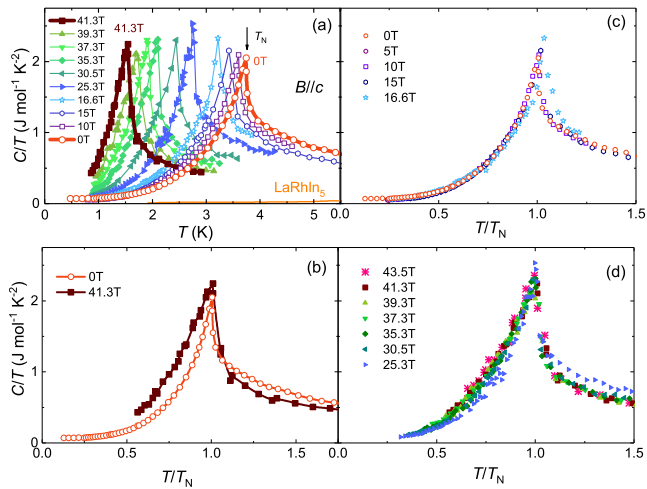}
\caption{(a) Temperature dependence of the specific heat $C$/$T$ of CeRhIn$_5$ in
various magnetic fields with $B$~$\parallel$~$c$. Data between 0--15 T were measured in a PPMS while
the high-field data was measured in pulsed magnetic fields. The orange curve at low $C/T$ is the
specific heat of LaRhIn$_5$ at $B$ = 0. (b) $C$/$T$ as a function of $T/T_N$ in zero field and 41.3~T, one
 field above and one below $B^*$. $C$/$T$ is displayed as a function of $T/T_N$  for applied fields (c) up to 16.6~T, and (d) at fields higher than 16.6~T.}
\label{Fig2}
\end{figure}

Figure~\ref{Fig2}(a) shows the temperature dependence of the specific heat of
CeRhIn$_5$ [$C(T)$] at different magnetic fields with $B$~$\parallel$~$c$, reaching a higher maximum field, and lower temperatures in the high-field region than previous studies of $C(T)$ \cite{Kim}; here the level of precision allows for a detailed study of the evolution of electronic correlations.
The AFM transition is clearly observed at all fields, which is
manifested by a pronounced peak in $C(T)/T$. With increasing magnetic
field, $T_N$ smoothly shifts to lower temperatures, consistent with previous
reports \cite{Jiao,Kim}. The accuracy of the current data obtained in pulsed
magnetic fields was checked by comparing to those measured in dc fields \cite{Kim,sup}. $C/T$ in the
displayed temperature range is much larger than in the La homologue, and therefore the phonon contribution is negligible.

Figure~\ref{Fig2}(b) displays $C/T$ as a function of $T/T_N$ for two representative curves [bold in Fig.~\ref{Fig2}(a)], at $B=0$ and 41.3~T, which show distinctly different behavior. At high fields, the transition is sharper, with a larger jump, while the values of $C/T$ just above $T_N$ are reduced. This suggests that compared to the low-field results, at high fields there is a reduction
in the contribution from short-range fluctuations which appear above the magnetic ordering temperature. In Fig.~\ref{Fig2}(c), $C/T$ is displayed for fields up
to 16.6~T as a function of $T/T_N$  \cite{sup}. One can see that the data below $T_N$ overlap fairly well, demonstrating that $T_N$
is the only energy scale which determines the specific heat in this field range. On the other hand, the higher field data do not
scale so well as a function of $T/T_N$, suggesting the role of additional parameters. This may also be taken as an indication of changes in the underlying correlated state, such as a variation of the effective mass, as described below.

We further analyzed the data well below $T_N$ using $C$/$T$ = $\gamma$ + $C_m$/$T$, where $\gamma$ is the Sommerfeld coefficient, and $C_m$ is
the contribution of the AFM magnons to the specific heat \cite{Continentino,sup}. This equation can be fitted well to
the low-temperature data [$T$ $<$ 0.7$T_N$]  below 35.3~T. Above 35.3~T,
the fit is less reliable due to the limited accessible data  for $T$ $<$ 0.7$T_N$. A
detailed discussion of the specific heat fits is presented in the Supplemental Material \cite{sup}, where with increasing magnetic field, $\gamma$($B$) changes little at lower fields,
but shows an enhancement onsetting at fields above  $B_{\rm onset}\approx17$~T.

\begin{figure}[tb]\centering
 \includegraphics[width=6cm]{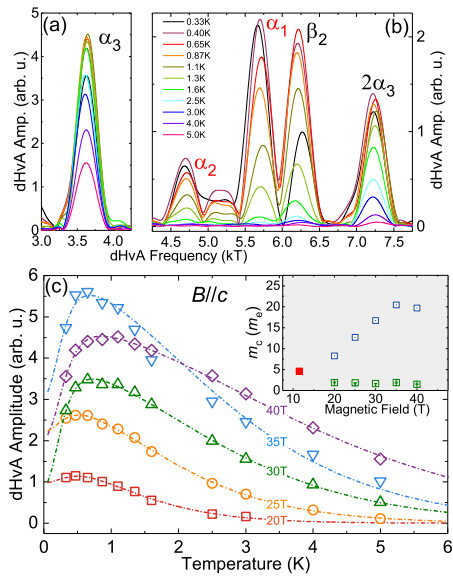}
\caption{(a) and (b) FFT spectra of dHvA effect oscillations of CeRhIn$_5$ for the field range
35--45 T at various temperatures with for $B \parallel c$. (c) Temperature dependence of dHvA oscillation
amplitudes of the $\alpha_3$ orbit at various magnetic fields. The field values correspond to the midpoints of the 10~T wide field windows used in the FFT. The dashed lines
display fits to the spin-dependent LK formula. The inset shows the field dependence
of the cyclotron masses of the two spin orientations (open blue and green squares).
Experimental data obtained at low magnetic fields from
Refs.~\onlinecite{Cornelius00,Hall} are also shown by the red solid squares.}
\label{Fig3}
\end{figure}

\subsection{Analysis of the effective cyclotron masses from the dHvA effect}

To directly probe the effective mass of the charge carriers, we also measured the dHvA effect.
It has been established by band-structure calculations and quantum-oscillation
measurements that the FS of CeRhIn$_5$ primarily arises from three bands ($\alpha$,$\beta$,$\gamma$), where the dHvA effect is dominated by the extremal orbits denoted $\alpha_1$, $\alpha_2$, $\alpha_3$, $\beta_1$, and
$\beta_2$ \cite{Hall,Shishido2002}. In our previous work
(Refs.~\onlinecite{Jiao,Jiao2}), we studied the magnetic field dependence of the dHvA
frequencies at 0.33 K, which yielded clear evidence for an abrupt expansion of the FS volume at $B^*$. Here, we further investigate the temperature and magnetic-field
dependence of the dHvA amplitudes, from which the effective cyclotron masses of the charge carriers are determined, focussing on the $\alpha_3$ orbit, which corresponds to the strongest oscillation.

In a conventional metal, the dHvA oscillation amplitudes are given by the
Lifshitz–Kosevich (LK) formula \cite{Shoenberg}, where in the nonspin-dependent case the temperature and field dependence are determined by two adjustable parameters, the effective cyclotron mass ($m_c$) and the Dingle temperature ($T_D$),  the analysis of the latter being presented in the Supplemental Material \cite{sup}. Figures~\ref{Fig3}(a) and (b) display  the fast Fourier transform (FFT) spectra of the dHvA oscillations of CeRhIn$_5$ at various
temperatures for 35~T~$\leq B \leq 45$~T with $B\parallel c$.
\begin{figure}[tb]\centering
 \includegraphics[width=7cm]{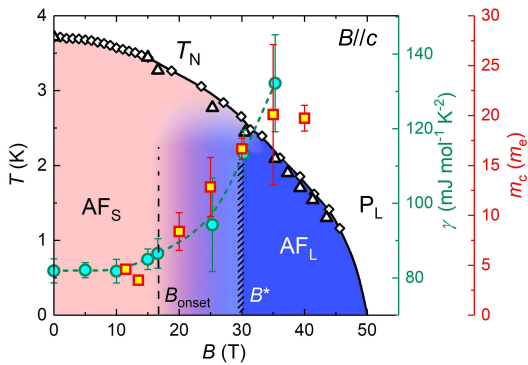}
\caption{Temperature-field phase diagram of CeRhIn$_5$ where the field dependence of $\gamma$ and the effective mass, $m_c$, of the heavy spin orientation of the FS sheet for $B \parallel c$ [see inset of Fig.~\ref{Fig3}(c)] are  shown by the olive circles and red squares respectively. The scales for these quantities with corresponding colors are on the right side of the panel.  The positions of the FS reconstruction ($B^*$) and the field at which the effective mass enhancement onsets ($B_{\rm onset}$) are also displayed. The triangular and diamond shaped symbols represent $T_N$ as obtained here and in Ref. \onlinecite{Jiao}, respectively.}
\label{Fig4}
\end{figure}
If the data were well described by the LK  formula, the amplitude of the oscillations
would monotonically increase with decreasing temperature. As shown in Fig.~\ref{Fig3},
this is the case for the $\alpha_3$ orbit above 1~K only, but below this temperature
the dHvA amplitude reaches a maximum before it decreases with decreasing temperature. This decrease cannot arise due to an increased scattering rate, since it is found that $T_D$ decreases strongly in this temperature and magnetic field range \cite{sup}. Such a phenomenon was previously observed in a
pulsed-field experiment in the range 31--50 T and was attributed to
the formation of SDW order \cite{Cornelius00}. However, this explanation cannot account for the similar  observations of the $\alpha_3$ orbit of CeCoIn$_5$, where it was ascribed to spin-dependent
mass enhancements of the FS \cite{McCollam}. This phenomenon may arise in heavy fermion systems when a sufficiently high magnetic field lifts the spin degeneracy of the renormalized bands by the Zeeman effect. Due to the asymmetry of the electronic bands near the Fermi level in the heavy fermion state, this can lead to  different effective masses for the two spin species \cite{McCollam}. Therefore, we studied the
spin-dependent effective masses of the electrons on the $\alpha_3$ orbit by performing
FFT on the dHvA oscillations at various temperatures in 10~T
intervals.

Following Refs.~\onlinecite{Takashita,McCollam}, a spin-dependent LK formula was
used to fit the derived amplitudes given by $\tilde{M}~=~\sqrt{(\tilde{M}_{\uparrow}+\tilde{M}_{\downarrow})^2\textrm{cos}^2
\theta+(\tilde{M}_{\uparrow}-\tilde{M}_{\downarrow})^2\textrm{sin}^2\theta}$, where $\theta$ is the phase term and $\tilde{M}_{\uparrow}$ ($\tilde{M}_{\downarrow}$)
is the dHvA amplitude of the spin up (down) FS described by the LK formula. This model fits the experimental results well [dashed-dotted
lines in Fig.~\ref{Fig3}(c)]. As shown in the inset of Fig.~\ref{Fig3}(c), the fit reveals a relatively large field-dependent effective mass
for one of the spin orientations but a small, nearly field-independent mass for the
other one. It should be noted that we cannot determine which spin orientation corresponds to the carriers with the larger effective
mass. The mass of the lighter band is around 1.6$m_e$ in the measured field
range, which is slightly larger than for the corresponding orbit in
LaRhIn$_5$~\cite{Shishido2002}, while the mass of the heavier band increases
with increasing magnetic field.  Note that this mass reaches a value of about 20$m_e$ when the midpoint field is 35~T, at which the dHvA amplitude displays a significantly steeper temperature dependence. Such a heavy mass at a field above $B^*$ is similar to that of the corresponding orbit of CeCoIn$_5$ \cite{McCollam}, where the $4f$ electrons also contribute to the Fermi surface \cite{Harrison2004}. A similar suppression of the dHvA amplitude at low temperatures is
observed for the $\beta_2$ orbit \cite{sup}. Since dHvA signals of the $\alpha_2$, $\alpha_1$, and $\beta_1$
orbits only appear below 2.5~K and above 30~T, the field dependences of these cyclotron masses cannot be estimated.

\section{Discussion}

Our main results are  summarized in Fig.~\ref{Fig4}, which show that although $T_N$ is smoothly suppressed as a function of applied field, there is an enhancement of the $\gamma$ coefficient and effective
cyclotron mass $m_c$ which onsets above  $B_{\rm onset}\approx17$~T, well below $B^*\approx30$~T. Combined with our previous results \cite{Jiao,Moll,Jiao2}, we are able to identify a number of distinct behaviors across different field ranges at low temperature. Between zero field and $B^*$, the $4f$ electrons are well localized and the Fermi surface is small. At lower fields up to $B_{\rm onset}$,  $\gamma$ barely changes with increasing field, and there is also little change in the in-plane magnetostriction. However, upon further increasing the field, the specific heat and dHvA measurements are consistent in revealing a clear enhancement of the effective quasiparticle mass, which likely signals an increasing fraction of the large FS in the fluctuation spectrum of the FS crossover \cite{JMMMXX}. We note that an anomaly at $B_M\approx18$~T was  previously found in Hall-resistivity measurements \cite{Jiao}, but whether this is related to the onset of the electronic mass enhancement requires further determination. At the subsequent crossover at $B^*$ $\approx$ 30~T, the Fermi surface becomes reconstructed from small to large.
Here, a maximum of the effective mass at $\approx$ 30~T may be anticipated, corresponding to the finite-temperature signature of a zero-temperature divergence  \cite{Gegenwart}. However, it is difficult to confirm whether such a maximum occurs, due to  the extremely large error bar of the cyclotron mass for the field-window centered around 35~T, in comparison to the preceding and following data points at 30~T and 40~T, respectively. It can be seen  in Fig.~\ref{Fig3}(c) that the 35~T data shows the steepest increase with decreasing temperature and is less well described by the  LK-formula for spin-dependent cyclotron masses, consistent with a rapid field-induced change of the electronic structure in this field-window. On the other hand, the absence of such a maximum  might be related to there being  a sizeable mass enhancement only for \textit{one} spin orientation. The high-field mass enhancement  significantly affects $C(T)$, where the analysis of the low-temperature data suggests a pronounced weakening of the RKKY coupling strength above $B_{\rm onset}$ \cite{sup}.  In addition, the short-ranged correlations above $T_N$ appear to be substantially reduced, although the magnetic structure likely remains unchanged to fields above $B^*$ \cite{UrbanoPoster}.

The field-induced increase of  the effective mass, together with  the observed changes to $C(T)$ in the magnetic state and the drastic changes of some dHvA frequencies upon crossing  $B^*$ with others remaining at a nearly constant value \cite{Jiao},  implies that the Fermi surface reconstruction is driven by the delocalization of the $4f$ electrons. Whether this delocalization transition occurs as an orbitally  selective Mott transition or other  forms of Kondo destruction is an interesting open question. Addressing it may require relating the observed changes of certain dHvA frequencies to details of the Fermi surface, and thus disentangling the effects of the $f$-electron degrees of freedom, momentum dependence of the hybridization and band structure of the conduction electrons. This may shed new light on how orbital and spin degrees of freedom are entangled in strong magnetic fields. Recent results indicating the emergence of an anisotropy of the electrical resistivity due to spin nematicity \cite{Ronning2017} at  $B$ $\approx$ $B^*$ in CeRhIn$_5$ also show a close relationship to the SDW phase with itinerant $4f$ electrons. Indeed, in the low-field regime no resistivity anisotropy is observed but at intermediate fields \textit{below} 30~T,  there is an onset of this resistivity anisotropy  near $B_{\rm onset}$ \cite{sup}.  This in turn suggests that the presence of itinerant, rather than localized, $4f$-electron states is related to the emergence of this new electronic (nematic) order.

\section{summary}

To conclude, we have probed the nature of the Fermi surface reconstruction associated with the  abrupt change of the Fermi surface volume of CeRhIn$_5$ at $B^*\approx30$ T. We find a field-induced enhancement of the effective mass which onsets  at  $B_{\rm onset}\approx17$~T and significantly increases with increasing field below $B^*$, as well as distinct changes to the temperature dependence of the specific heat.  These new insights into the Fermi surface  reconstruction which takes place inside the magnetically ordered states of the Kondo-lattice system CeRhIn$_5$ should  be relevant to other heavy-fermion compounds, as well as a broader range of correlated metals in proximity to Mott insulating states, such as high-$T_c$ cuprates or organic charge-transfer salts.

\begin{acknowledgments}
We acknowledge valuable discussions with S.~Arsenijevic, P.~J.~W.~Moll, Q.~Si, F.~Ronning, S.~Paschen, O.~Stockert and C.-L.~Huang.
Work at Zhejiang University was supported by the National Natural Science Foundation of China (Grants No. U1632275 and No. 11474250), National Key R\&D Program of China (Grants No. 2017YFA0303100 and No. 2016YFA0300202), and the Science Challenge Project of China (No.~TZ2016004). Work at Los
Alamos National Laboratory was performed under the auspices of the US
Department of Energy (DoE). A portion of this work was performed at the National High Magnetic Field Laboratory, which is funded by the National Science Foundation through
DMR-1157490 and the US Department of Energy and the State of Florida. J. S. appreciates funding from Basic Energy Sciences, U. S. Department of Energy FWP {\it Science in 100 T} program. We acknowledge the support of the HLD at HZDR, member
of the European Magnetic Field Laboratory (EMFL), and support by the ANR-DFG grant Fermi-NESt.
L.J. acknowledges support by the Alexander-von-Humboldt foundation.
\end{acknowledgments}

\end{document}